\centerline
{\bf Should the Strange Magnetic Moment of the Nucleon be Positive?} 
\medskip
\centerline
{R. D. McKeown}
\centerline
{\it W. K. Kellogg Radiation Laboratory}
\centerline
{\it California Institute of Technology}
\centerline
{Pasadena, CA 91125}
\bigskip
\centerline
{ABSTRACT}
{\narrower \smallskip The strange magnetic moment of the nucleon ($\mu_s$) is
examined as part of the nucleon's isoscalar anomalous moment. The
dominant up and down  quark effects in the anomalous moment may
actually tend to favor
$\mu_s >0$, which is contrary to the negative values that generally 
result from model calculations. The possible origins of this apparent
discrepancy are considered.
 \smallskip }

\bigskip

Several years ago, Kaplan and Manohar${}^1$ pointed out that the neutral weak
current could be used to determine the strange quark-antiquark 
($\bar s s$) contributions
to nucleon form factors. These $\bar s s$ effects are very interesting
in that they represent low energy properties of the nucleon which must
be due to the quark-antiquark sea. Determining such quantities
can thus provide deeper insight into the origins of nucleon properties in terms
of QCD.

The magnetic moment is an excellent example of a nucleon property that
can be studied in this fashion. 
Electromagnetic couplings probe only the charged constituents of the nucleon:
the quarks and antiquarks.
The anomalous components of the nucleon
magnetic moments are known to be large, but we still do not
have a satisfactory explanation of
their magnitudes in the context of QCD. While many models
do give reasonable values, none are firmly based 
(without additional assumptions) on QCD.
Thus, the real origin of the
nucleon's anomalous magnetic moment in terms of QCD 
(rather than ad hoc models like constituent quarks) has been elusive.

After the suggestion of Kaplan and Manohar, it was proposed that the
neutral weak magnetic moment of the proton could be measured in parity
violating electron scattering.${}^2$ Indeed, such an experiment 
(the SAMPLE experiment${}^3$) is currently
in progress and results are expected soon.

In this Letter I examine the role of $\mu_s$, the strange quark-antiquark 
contribution
to the nucleon's isoscalar anomalous moment. In particular, it will be
suggested that it may be very natural to expect that the value of $\mu_s$
is positive, rather than the generally negative values obtained in
model calculations.${}^{4-9}$

In general, we can define the vector matrix elements corresponding to
each quark flavor as follows${}^{10}$:
$$ <N|\bar q_f \gamma^\mu q_f | N> \, \equiv  \, \bar u_N \bigg[ F_1^f(k^2) 
\gamma^\mu
+  {i \over {2M_N}}  F_2^f(k^2)  
\sigma^{\mu \nu} k_\nu \bigg] u_N \eqno (1)
$$
in which $q_f$ is a quark field operator of flavor $f = u, d, s$, and the
$F_1^f$ and $F_2^f$ are form factors which are functions of the squared
momentum transfer. The electromagnetic nucleon form factors are then obtained
as functions of the individual flavor form factors:
$$
F_{1,2} = {2 \over 3} F_{1,2}^u - {1 \over 3} F_{1,2}^d - {1 \over 3} F_{1,2}^s
\eqno (2) 
$$
where the coefficients represent the electromagnetic couplings to the
individual quarks (i.e., the electric charges).
Similarly, the neutral weak couplings can be written 
$$
F_{1,2}^Z = ({1 \over 4} - {2 \over 3} \sin^2 \theta_W)  F_{1,2}^u
		+(- {1 \over 4} + {1 \over 3} \sin^2 \theta_W)  F_{1,2}^d
		+(- {1 \over 4} + {1 \over 3} \sin^2 \theta_W)  F_{1,2}^s \>.
\eqno (3)
$$
(Radiative corrections slightly modify this tree level expression.${}^{11}$)
Thus we can see that
a determination of the isoscalar, isovector, and neutral weak form factors 
would allow solving for the contributions of all 3 flavors.

In the following, we will explore the use of eq. (2) 
at $k^2 = 0$ and the implications of
the well-known electromagnetic moment values  for the flavor components
of the anomalous moments. Beginning first with $F_1$ (equal to the 
nucleon electric charge at $k^2=0$),
we find 
$$
\eqalign 
{
Q_p &= {2 \over 3} F_1^u - {1 \over 3} F_1^d   \cr
Q_n &= {2 \over 3} F_1^d - {1 \over 3} F_1^u   \cr} \eqno (4)
$$
where we have assumed isospin symmetry and define 
$F_1^u \equiv F_1^u$(proton)$= F_1^d$(neutron), 
$F_1^d \equiv F_1^d$(proton)$= F_1^u$(neutron).
Note also that we have set $F_1^s=0$ since there is no
net contribution of strange quarks to the nucleon charge. (This is {\it not}
true for the magnetic moment.)

Using eq. (4) we obtain the trivial results $F_1^u = 2$ and $F_1^d = 1$, 
which of
course are just the valence numbers of the quarks. Now let's apply eq. (2)
to the anomalous moments. Then we obtain
$$
\eqalign {
F_2^p &= {2 \over 3} F_{2}^u - {1 \over 3} F_{2}^d  - {1 \over 3} \mu_s 
 \cr
F_2^n &= {2 \over 3} F_{2}^d - {1 \over 3} F_{2}^u  - {1 \over 3} \mu_s \>. 
 \cr}\eqno (5)
$$
Using the empirical values of $F_2^p = 1.79$ and $F_2^n = -1.91$, and forming
isovector and isoscalar combinations, we get
$$
\eqalign 
{
3.70 &= F_2^u-F_2^d \cr
-0.12 &= {1 \over 3} (F_2^u + F_2^d) - {2 \over 3} \mu_s  \>.  \cr} \eqno (6)
$$
We have 2 equations and 3 unknowns. Determination of $\mu_s$ would then
allow solving for the other 2 flavor form factors. 

Nevertheless, we can gain
some insight from eq. (6). First, one should note that the left side of the
isoscalar equation is small compared to the isovector. Since we do not
expect $\mu_s$ to be large (generally $|\mu_s| <1$), we see that roughly
$F_2^u \approx - F_2^d \sim 2$. It is striking that these anomalous moments
associated with the up and down quark flavors are so large but similar
in absolute magnitude. From eq. (6) we see that the relative magnitudes
of $F_2^u$ and $F_2^d$ are closely related to the sign of $\mu_s$
So the next question we might ask is whether
we expect $|F_2^u| > |F_2^d|$ or vice versa. 

Our knowledge of the flavor dependence of other nucleon properties
indicates that the up quark effects are generally dominant. 
Valence quantities clearly show this effect due to the fact that
there are 2 
valence up quarks and only one valence down quark.
For example, this is
manifestly true in the valence quark counting involved with
$F_1^u = 2 F_1^d$. Another indication comes from the axial matrix elements, 
$$
\Delta q_f = - <N | \bar q_f \gamma^\mu \gamma^5 q_f | N > S_\mu \eqno (7)
$$
where $S_\mu$ is the nucleon spin vector.
These matrix elements are probably more relevant to the
$F_2^f$ since they also involve the
spin couplings of the quarks. 
Recently, spin dependent
deep inelastic scattering experiments${}^{12}$ have determined 
$\Delta u = 0.85 \pm 0.03$  and 
$\Delta d = -0.41 \pm 0.03$, so the spins of the up quarks are 
dominant. 
If $F_2^u$ and $F_2^d$ have similar behavior, then we would have
$|F_2^u| > |F_2^d|$. 

One area where it has been found that the up quarks do not dominate
is in the antiquark sea. Measurements of the Gottfried sum rule indicate 
that${}^{13}$
$$
\int_0^1 dx [\bar u (x) - \bar d (x)] = -0.147 \pm 0.039 \>. \eqno (8)
$$ 
Here $\bar u (x)$ and $\bar d (x)$ are deep inelastic structure functions
corresponding to the antiquarks, and are functions of the momentum 
fraction $x$.
However, this is a pure sea quantity with no valence contribution at all.
In addition, there is no spin dependence in these quantities.
It would seem that the dominance of the up quarks in the axial matrix elements
is more relevant to the magnetic moment discussion.

Regarding the magnetic moments,
there is no apriori requirement that the compositeness of the nucleon
should generate large anomalous magnetic moments (i.e., $F_2^u=F_2^d=0$
would certainly be allowed in principle).
Clearly, the quantities $F_2^u$ and $F_2^d$ are generated dynamically
and this is a very large (not subtle) effect. Thus it would seem that
we should expect these quantities to display the dominance of up
quarks that we discussed above, leading to $|F_2^u| > |F_2^d|$.

Now we return to eq. (6). If $|F_2^u| > |F_2^d|$ holds, then the 
isoscalar equation implies that $\mu_s>0$. In fact, if $\mu_s<0$ then
one is forced to conclude that $|F_2^d| > |F_2^u|$. So we see that
it actually seems more natural to expect $\mu_s >0$.

Table 1 shows the values of $F_2^u$ and $F_2^d$ for some typical values of
$\mu_s$. Note that a negative value of $\mu_s$ causes a reversal 
in the relative magnitudes of $F_2^u$ and $F_2^d$, which is 
contrary to the expectations based on the above discussion.
Therefore, it seems that
a positive value of $\mu_s$ is a rather natural expectation based on these
considerations.

\midinsert
\centerline
{\bf Table 1}

$$\vbox{\offinterlineskip
\halign{\strut\quad\hfil#& \vrule#& \quad#\hfil& \vrule#& \quad#\hfil\cr
$\mu_s$	     &&	$F_2^u$	&&	$F_2^d$\cr
\noalign{\vskip 2pt}
\noalign{\hrule}
\noalign{\vskip 1pt}
$-{1 \over 2}$ &&	1.17	&&	-2.53\cr
0	     &&	1.67	&&	-2.03\cr	
${1 \over 2}$  &&	2.17	&&	-1.53\cr
}}$$
\endinsert

In general, model calculations${}^{4-9}$ of $\mu_s$ give $\mu_s<0$. 
(I do note that one reference${}^{14}$ does predict a positive 
$\mu_s = 0.42 \pm 0.3$
based on an SU(3) chiral hyperbag model.)
These models typically
invoke a specific mechanism, such as the fluctuation 
of a proton into
a $\Lambda$ and $K^+$, for generating $\mu_s$. 
However, such dynamical mechanisms are probably only 
part of the total strange magnetic moment. For example, the constituent quarks
in the baryons may have intrinsic $\bar s s$ pairs which are not dynamical
degrees of freedom in baryon-meson or constituent quark models. Perhaps there
is a deeper reason why the constituent quarks themselves may make a 
positive contribution to $\mu_s$. Clearly, the experimental determination of
this quantity will be extremely interesting and will provide
an important new clue to the structure of the nucleon.

This work was supported by the National Science Foundation (PHY-94204720).

\medskip
\noindent
{\bf References}

\noindent
${}^1$ D. Kaplan and A. Manohar, Nucl. Phys. B {\bf 310}, 527 (1988).

\noindent
${}^2$ R. D. McKeown, Phys. Lett. {\bf B219}, 140 (1989).

\noindent
${}^3$ Bates proposal 89-06, R. D. McKeown and D. H. Beck, spokespersons.

\noindent
${}^4$ R. L. Jaffe, Phys. Lett. {\bf B229}, 275 (1989).

\noindent
${}^5$ M. J. Musolf and M. Burkardt, Z. Phys. {\bf C61}, 433 (1994).

\noindent
${}^6$ W. Koepf, E. M. Henley, and S. J. Pollock, Phys. Lett {\bf B288} 
11 (1992).

\noindent
${}^7$ N. W. Park, J. Schechter, and H. Weigel, Phys. Rev. {\bf D43}, 
869 (1991).

\noindent
${}^8$ S. C. Phatak and Sarira Sahu, Phys. Lett {\bf B321} 11 (1994).

\noindent
${}^9$ Chr. V. Christov, {\it et al.,} Prog. Part. Nucl. Phys. {\bf 37}, 1
(1996).

\noindent
${}^{10}$ D. H Beck, Phys. Rev. {\bf D39}, 3248 (1989).

\noindent
${}^{11}$ M. J. Musolf and B. R. Holstein, Phys. Lett. {\bf B242}, 461 (1990).

\noindent
${}^{12}$ J. Ellis and M. Karliner Phys. Lett. {\bf B341} 397 (1995).

\noindent
${}^{13}$ P. Amaudrux {\it et al.,} Phys. Rev. Lett. {\bf 66} 2712 (1992).

\noindent
${}^{14}$ S. Hong and B. Park, Nucl. Phys. {\bf A561}, 525 (1993).

\end

\noindent
${}^6$ R. S. VanDyck, D. L. Farnham, and P. B. Schwinberg, Phys. Rev. Lett.
{\bf 70}, 2888 (1993).

\end